\documentclass{article}
\begin{document}
\begin{center}
{\Large \bf {Dark matter in the framework of shell-universe}
}\\
\vskip 0.5cm
{Merab Gogberashvili} \\
\vskip 0.3cm {\it
{Andronikashvili Institute of Physics}\\
{6 Tamarashvili Str., Tbilisi 0177, Georgia}\\
{(E-mail: gogber@hotmail.com)}} \\
\vskip 0.3cm
{Michael Maziashvili} \\
\vskip 0.3cm {\it
{Department of Theoretical Physics, Tbilisi State University \\
3 Chavchavadze Ave., Tbilisi 0128, Georgia} \\
{(E-mail: maziashvili@ictsu.tsu.edu.ge)}} \vskip 0.5cm
{\Large \bf Abstract}\\
\quotation

{\small We show that the shell-universe model (used to explain the
observed expansion rate of the universe without a dark energy
component) provides a natural mechanism for local increasing of
the brane tension leading to the modified Newton's law alternative
to galactic dark matter.

\vskip 0.2cm \noindent Kay words: Brane, Dark matter, MOND.  }
\endquotation \end{center}

\section{Introduction}
Numerous astrophysical observations have shown that classical
Newtonian dynamics fails on galactic scale and beyond, if only
visible matter is taken into account. So long the only evidence
for the dark matter is its global gravitational effect. The two
most popular theoretical concepts dealing with this problem are
dark matter (DM) and modified Newtonian dynamics (MOND). In other
words, either the universe contains large quantities of unseen
matter, or gravity is not generally the same as it appears to be
in the solar system.

MOND, alternative to cosmic dark matter was proposed in \cite{Mil}
to explain flat rotation curves or the existence of a
mass-rotation velocity relation for spiral galaxies. An obvious
first choice would be to propose that gravitational attraction
force becomes more like $1/r$ beyond some length scale, which is
comparable to the scale of galaxies \cite{San}. In this model a
test particle at a distance $r$ from a large mass $M$ is subject
to the acceleration
\begin{equation}\label{milgmod}
|\vec{a}| = \frac{G_NM}{r^2}g(r/r_0)~,
\end{equation}
where $G_N$ is the Newtonian constant, $r_0$ is of the order of
the sizes of galaxies (a few kpc) and $g(r/r_0)$ is a function
with the asymptotic behavior \cite{San}:
\begin{eqnarray}
g(r/r_0)=1~,~~~~~(r \ll r_0)~, \nonumber \\
g(r/r_0)=r/r_0~, ~~~~~ (r \gg r_0)~ .
\end{eqnarray}

The fact that, to some extant, both DM and MOND can successfully
explain galactic dynamics favors the possibility that there exists
a deeper connection between these two theories \cite{Dunk}. Let us
consider one such possibilities in the braneworld approach.

Throughout this paper we assume that galaxies containing only the
observable baryonic matter are a fair sample of the universe. We
want to consider the model of the universe as a 3d spherical brane
expanding in a 5d space-time. This brane provides the localization
of matter on the $S^3$ and in a large-scale approximation can be
regarded as a homogenous and isotropic closed universe. The
shell-universe model allows one to explain the observed expansion
rate of the universe without a dark energy component \cite{Gog1}.
Two observed facts of cosmology, the isotropic runaway of galaxies
and the existence of a preferred frame in Universe where the relic
background radiation is isotropic, also find their obvious
explanation in shell-universe model.

Here we consider the shell-universe model on the scale
characterized by the galaxy size and study the possibility how the
galaxy in this model can modify Newton's law at large distances.
The discussion is present in section $2$. Section $3$ is devoted
to the summary and in the appendix we recall some results from the
theory of elasticity.

\section{Galaxy in/on the shell-universe}

We shall consider the shell-universe model with the negative
cosmological constant, which is the same inside and outside of the
shell \cite{Gog1}. Assuming that the present value of the shell
radius much exceeds the galaxy size one can approximate the shell
locally (at the distances comparable to the galaxy size) by the
well known flat brane solution \cite{Gog2, RS1}. The flat brane
solution embedded in $AdS_{5}$ and located at $y=y_0$ has the form
\begin{equation}\label{grs}
ds^2=e^{-2\varepsilon|y-y_0|}\eta_{\mu\nu}dx^{\mu}dx^{\nu}+dy^2~,
\end{equation}
where $\mu,~\nu = 0,~1,~2,~3$, parameter $y_0$ denotes the present
value of the shell-universe radius,
$\varepsilon=\sqrt{-\Lambda/6}$ characterizes the width of the
brane and $\eta_{\mu\nu}= diag(-\,+\,+\,+)$ is the four
dimensional Minkowskian metric. The negative bulk cosmological
constant $\Lambda$ is adjusted as $\kappa^4\lambda^2/6= -\Lambda$,
where $\lambda$ is the brane tension and $G=\kappa^2/8\pi$ is the
5d gravitational constant

Now let us put the galaxy on the brane. In the presence of matter
on the brane (\ref{grs}), the linearized approximation to the
gravitational field requires choosing a gauge in which the brane
is bent in some neighborhood of the matter \cite{GT,GKR}. However,
as it was shown a bit later this kind of brane bending is the
artefact of incorrect gauge fixing condition \cite{AIMVV} and
should be ruled out. Nevertheless, it is of interest to view what
kind of modification of Newton's law arises due to this bending,
which we parameterize as $y=y_0-\xi(x^{\mu})$.

The real bending of the brane $\delta\xi$ appears  because of the
gravitational attraction the galaxy by the shell-universe. The
galaxy deforms some neighborhood of that part of the brane upon
which it acts and leads to the appearance of an "elastic" force.
We have introduced the "elasticity" of the brane into play in
order to compensate the gravitational force between the galaxy and
the shell-universe. Hence, in the region the galaxy is situated
the energy density of the brane (i.e. the brane tension $\lambda$)
is increased. Generally speaking, the increment of tension is
proportional to the gravitational force by which the
shell-universe attracts the galaxy and is inversely proportional
to the galaxy volume. Since the core of the galaxy is much more
dense than its outer part, it is natural to assume that the
deformation of the brane is defined mainly by the galaxy core and
thus with a good accuracy is spherically symmetric. By taking into
account that the mass of the shell-universe is $2\pi^2\lambda
y^3_0$ and in 5d space-time the gravitational force obeys the
inverse-cube law one finds that the magnitude of gravitational
force which acts on the unit of brane area is proportional to
$\kappa^2\lambda T_{00}(\vec{x})$, where $\vec{x}$ are spatial
coordinates on the brane and $T_{00}(\vec{x})$ denotes the matter
density of the galaxy. Due to its "elastic" nature the increment
of the brane tension, i.e. the deformation energy density, is
equal to
\begin{equation}\label{defenerg}
\delta\lambda=\frac{D}{2}(\triangle\delta\xi)^2~,
\end{equation}
where $\triangle$ denotes the 3d Laplace operator and $D$ is the
"stiffness" coefficient of the brane \cite{LL} (see Appendix).
Using the equation describing the deformation of an elastic plate
one finds that $\delta\xi$ is governed by
\begin{equation}\label{brdeform}
\triangle^2\delta\xi=\frac{\pi\kappa^2\lambda}{4D}T_{00}(\vec{x})~,
\end{equation}
with the boundary conditions
\begin{equation}
\delta\xi(R)=0~,~~~~~\triangle\delta\xi(R)=0~, \nonumber
\end{equation}
where $R$ is the radius of concavity.

Far from the center of galaxy ($r\gg r_0$) the eq.(\ref{brdeform})
reduces to
\begin{equation}\label{fargcenter}
\triangle^2\delta\xi=\frac{\pi\kappa^2\lambda
M}{4D}\delta(\vec{x})~,
\end{equation}
where $M$ denotes the mass of galaxy. From eqs.(\ref{defenerg})
and (\ref{fargcenter}) one finds
\begin{eqnarray}\label{labehav}
\delta\xi=-b\frac{\left(R-r\right)\left(2R-r\right)}{3R}~, \nonumber \\
\delta\lambda=2b^2D\left(\frac{1}{R}-\frac{1}{r}\right)^2 ,
\end{eqnarray}
where
\begin{equation} \label{b}
b=-\frac{\kappa^2\lambda M}{32D} = -\frac{3\epsilon M}{16D}~.
\end{equation}

From now on we assume that the coordinates $y,~\vec{x}$ are
Gaussian (normal) ones in some neighborhood of the 3-brane. The
gravitational equations on the $3$-brane take the form \cite{TKS}
\begin{equation}\label{efgreq}
G_{\mu\nu}=-\Lambda_4
g_{\mu\nu}+\frac{\kappa^4\lambda}{6}T_{\mu\nu}+
\kappa^4\Pi_{\mu\nu}-E_{\mu\nu}~,
\end{equation}
where $T_{\mu\nu}$ is the matter energy momentum tensor,
\begin{equation}
\Lambda_4=\frac{1}{2}\left(\Lambda+\frac{\kappa^4\lambda^2}{6}\right)
\nonumber
\end{equation}
is 4d cosmological constant, $E_{\mu\nu}$ is limiting value of the
electric part of the bulk Weyl tensor on the brane and
\begin{equation}
\Pi_{\mu\nu}=-\frac{1}{4}T_{\mu\alpha}T_{\nu}^{~\alpha}+
\frac{1}{12}TT_{\mu\nu}+
\frac{1}{8}g_{\mu\nu}T_{\alpha\beta}T^{\alpha\beta}-
\frac{1}{24}g_{\mu\nu}T^2 ~,
\end{equation}
here  $T=T_{\mu}^{~\mu}$.

In the low energy limit the quantities $E_{\mu\nu}$ and
$\Pi_{\mu\nu}$ are negligible and eq.(\ref{efgreq}) reduces to the
4d conventional Einstein gravity \cite{TKS}
\begin{equation}\label{congrav}
G_{\mu\nu}\simeq-\Lambda_4g_{\mu\nu}+\frac{\kappa^4\lambda}{6}T_{\mu\nu}~.
\end{equation}
Assuming $\delta\lambda(r)\ll\lambda$, in the Newtonian
approximation \cite{MTW}
\begin{equation}
T_{00}\gg\left|T_{\alpha i}\right|~ ,~~~~~ g_{00}=-1-2\varphi~,
~~~~~R_{00}=\triangle\varphi~, \nonumber
\end{equation}
the eq.(\ref{congrav}) takes the form
\begin{equation}\label{modnew}
\triangle\varphi=4\pi G_N\left[\delta\lambda(r)+T_{00}\right]~,
\end{equation}
where $G_N=\kappa^4\lambda/48\pi$.

From eq.(\ref{modnew}) follows that the potential about a mass $M$
has the form
\begin{equation}\label{modpot}
\varphi=-G_N\frac{M}{r}- G_N\int
d^3x'\frac{\delta\lambda(r')}{\left|\vec{r} - \vec{r}'\right|}~.
\end{equation}
From eq.(\ref{labehav}) one obtains that in the region $r_0\ll
r\ll R$ the tension increment behaves as
\begin{equation}
\delta\lambda\sim r^{-2} ~. \nonumber
\end{equation}
This leads to the appearance of a logarithmic potential far from
the galaxy center. More precisely, for the tension increment
$\delta\lambda$ given by eq.(\ref{labehav}) the solution of
eq.(\ref{modnew}) in the region $r_0\ll r < R$ behaves as
\begin{equation}\label{potfargalax}
\varphi=8\pi b^2DG_N\left\{\frac{r^2}{6R^2}-
\frac{r}{R}+\ln(r/\beta)\right\}-G_N\frac{M}{r}~,
\end{equation}
where $\beta$ is an integration constant with units of length.
Thus, the gravitational acceleration of test particle in the
region $r_0\ll r\ll R$ has the form
\begin{equation}
\vec{a}=-\vec{\nabla}\varphi=-\left(G_N\frac{M}{r^2}+ \frac{8\pi
b^2DG_N}{r}\right)\frac{\vec{r}}{r}~. \nonumber
\end{equation}
The compatibility of this attraction law with the
eq.(\ref{milgmod}) leads to the following fine-tuning condition
\begin{equation}
8\pi b^2D=\frac{M}{r_0} ~.\nonumber
\end{equation}
Substituting the expression (\ref{b}) this condition takes the
form
\begin{equation}\label{stiffcoeff}
r_0M=\frac{32D}{9\pi\epsilon^2}~.
\end{equation}
From this relation it follows that if the brane-width $\epsilon$
and the stiffness coefficient $D$ are universal parameters on the
brane the length scale $r_0$ (at which the deviation from Newton's
law becomes essential) is inversely proportional to the mass of
galaxy $M$. For the object with mass $\sim M_{\odot}\approx
10^{-11}M$ one finds that modification of Newton's law takes place
out of horizon and is unobservable.

Now let us consider the second possibility of modification of the
Newton's law related directly with the displacement of the brane
caused by the gravitational interaction between shell-universe and
the galaxy. In the RS gauge, in presence of matter on the brane,
the brane is bent with respect to a coordinate system based on the
flat brane. In presence of matter $T_{\mu\nu}$, the location of
brane is given by $y=y_0-\xi(x^{\mu})$, where $\xi(x^{\mu})$ is
the solution of the equation \cite{GT,GKR}
\begin{equation} \label{11}
\partial_{\mu}\partial^{\mu}\xi(x^{\mu})=\frac{\kappa^2}{6}T ~.
\end{equation}
An additional slice deformation $\delta\xi$ results the appearance
of a dark matter on the brane with the following equation
\begin{equation}\label{wrongform}
\triangle\delta\xi=\frac{\kappa^2}{6}(3p-\rho)~,\nonumber
\end{equation}
which comes from (\ref{11}), where $p$ and $\rho$ - dark matter
pressure and density are assumed to be time independent
quantities.

Using the eq.(\ref{brdeform}) one finds the following relation
between the dark and baryonic matters
\begin{equation}\label{darkbaryonic}
\triangle(3p-\rho)=\frac{3\pi\lambda}{2D}T_{00}~.
\end{equation}
Under assumption that dark matter is essentially pressureless the
eq.(\ref{darkbaryonic}) far from the galaxy takes the form
\begin{equation}
\triangle\rho =-\frac{3\pi\lambda M }{2D}\delta(\vec{x})~.
\end{equation}
Its solution with the boundary condition $\rho(r=R)=0$ has the
form
\begin{equation}
\rho=\frac{3\lambda M }{8D}\left(\frac{1}{r}-\frac{1}{R}\right)~,
\end{equation}
and is not consistent with the logarithmic modification of the
potential at large distances from the galaxy center. However, as
it was already mentioned, this type of dark matter disappears in
the correct gauge \cite{AIMVV}.

\section{Summary}

We have considered two possibilities related with the local
deformation of a brane which could account for the dark matter in
the galactic halo for 4d observer. Both types of dark matter
considered here are undetectable directly from the standpoint of a
4d observer, but they lead to the modification of Newton's law.

The first possibility related with the local increment of the
brane tension leads to the modification of Newton's law at large
distances $r\gg r_0$, which is in agreement with the
eq.(\ref{milgmod}). So, in our approach each massive object is
surrounded by the elastic energy halo, which effectively plays the
role of dark matter. Consider the brane as a vacuum configuration
the appearance of elastic energy is some kind of vacuum
polarization effect due to matter.

We notice that the present scenario may also be viable for the RS
model containing two branes \cite{RS1}. In this case the hidden
brane attracts or repulses the galaxy and thereby leads to the
deformation of visible brane in the vicinity of the region where
the galaxy is situated.

The second possibility, which is related directly to the bending
of a brane due to presence of a matter, does not lead to the
desirable modification of the Newton's law at large distances
(\ref{milgmod}). One needs not worry about this because, as it was
mentioned above, this kind of bending is an artefact of incorrect
gauge fixing condition \cite{AIMVV}, and thereby this possibility
is merely ruled out.

\medskip
{\bf Acknowledgments}
\medskip

The authors are greatly indebted to V. Bochorishvili, A. Sagaradze
and G. Skhirtladze for help during the work on this paper and to
I. Lomidze for useful conversations.

\section*{Appendix}

Following to the textbook \cite{LL}, we briefly recall the results
related with the small deformation of an elastic plate slightly
adapted to the 4-dimensional case. Introducing the displacement
vector $u^i(x^k)$ ($i,~k = 1,~2,~3$), which characterizes the
shift of the particles of an elastic body under the deformation,
the deformation tensor for small deformations can be written in
the form
\begin{equation}
u_{ik}=\frac{1}{2}\left(\partial_i u_k+\partial_ku_i\right)~.
\nonumber
\end{equation}
The strength tensor corresponding to this deformation is given by
\begin{equation}\label{strength}
\sigma_{ik}=\frac{E}{1+\sigma}\left[u_{ik}+\frac{\sigma}{1-
3\sigma}\delta_{ik}u \right],
\end{equation}
where $u=u_m^{~\,m}$, $E$ is Young's modulus and the Poisson's
coefficient $\sigma$ takes values in the interval
$-1\leq\sigma\leq 1/3$. The deformation energy density reads
\begin{equation}\label{defener}
{\cal E}=\frac{\sigma^{ik}u_{ik}}{2} =
\frac{E}{1+\sigma}\left[u^{ik}u_{ik}+\frac{\sigma}{1-3\sigma}u^2\right].
\end{equation}

Consider an elastic 3-dimensional plate located at $y=0$. Usually
the internal stresses appearing under the small deformation are
much greater than the force that acts on the unit of plate area.
This leads to the condition $\sigma_{ik}n^k=0$, where $n^k$ is
unit normal to the plate. By virtue of eq.(\ref{strength}) this
condition gives
\begin{equation}\label{surfcondit}
u_{1y}=u_{2y}=u_{3y}=0,~~~~~u_{yy}=\frac{\sigma
(u_{11}+u_{22}+u_{33})}{2\sigma-1}~.
\end{equation}
If displacement $u^y=-\zeta (x^1,~x^2,~x^3)$ is known then from
eq.(\ref{surfcondit}) one can find
\begin{equation}
u_1=y\,\partial_1\zeta~,~~~~~ u_2=y\,\partial_2\zeta~,~~~~~
u_3=y\,\partial_3\zeta~,\nonumber
\end{equation}
and correspondingly
\begin{equation}
u_{11}=y\partial^2_1\zeta~,~~~~~u_{22}=y\partial^2_2\zeta~,
~~~~~u_{33}=y\partial^2_3\zeta~.\nonumber
\end{equation}

To simplify our consideration we suppose $\sigma$ is close to
$1/3$. Under this assumption the second term in eq.(\ref{defener})
becomes dominant and by taking into account that
$u_{yy}=-y\triangle\zeta$ the deformation energy density of an
elastic plate, after integrating across the plate, takes the form
\begin{equation}
{\cal E}=\frac{D}{2}(\triangle\zeta)^2~,
\end{equation}
where $\triangle$ denotes the 3d Laplace operator and $D$ is the
stiffness coefficient. For the variation one obtains
\begin{eqnarray}\label{defenervariat} \delta\frac{1}{2}\int
d^3x(\triangle\zeta)^2=\nonumber\\=\int
d^3x\delta\zeta\triangle^2\zeta-\oint
df^k\left(\delta\zeta\partial_k\triangle\zeta-
\triangle\zeta\partial_k\delta\zeta\right)~,
\end{eqnarray}
where $df^k$ is the surface element directed along the exterior
normal. The equation governing the deformation is obtained by
equating $\delta{\cal E}$ to the variation of the potential energy
of plate. If $P$, the force per unit area of plate, is
perpendicular to the plate then by taking into account
eq.(\ref{defenervariat}) the equation governing deformation takes
the form
\begin{equation}
D\triangle^2\zeta=P~,\nonumber
\end{equation}
with the conditions
\begin{equation}
\delta\zeta=\triangle\zeta=0 \nonumber
\end{equation}
at the boundary surface.


\end{document}